\documentclass[conf]{new-aiaa}

\usepackage[utf8]{inputenc}
\usepackage{geometry}
\usepackage{graphicx}
\usepackage{todonotes}
\usepackage{amsmath}
\usepackage[version=4]{mhchem}
\usepackage{longtable,tabularx}
\begin{document}

\title{A semi-structured approach to curvilinear mesh generation around streamlined bodies}

\author{Julian Marcon\footnote{PhD Candidate, Department of Aeronautics. AIAA Member.}, Joaquim Peir\'{o}\footnote{Reader, Department of Aeronautics.}}
\affil{Imperial College London, South Kensington Campus, London SW7 2AZ, United Kingdom}

\author{David Moxey\footnote{Lecturer, College of Engineering, Mathematics and Physical Sciences}}
\affil{University of Exeter, Streatham Campus, Exeter EX4 4QF, United Kingdom}

\author{Nico Bergemann\footnote{Software Developer}, Henry Bucklow\footnote{Product Manager. AIAA Member.}, Mark Gammon\footnote{Technical Director and CADfix Product Manager. AIAA Member.}}
\affil{ITI --- International TechneGroup Limited, Cambridge, United Kingdom}

\date{\today}

\maketitle

\begin{abstract}

We present an approach for robust high-order mesh generation specially tailored to streamlined bodies.
The method is based on a semi-sructured approach which combines the high quality of structured meshes in the near-field with the flexibility of unstructured meshes in the far-field.
We utilise medial axis technology to robustly partition the near-field into blocks which can be meshed coarsely with a linear swept mesher.
A high-order mesh of the near-field is then generated and split using an isoparametric approach which allows us to obtain highly stretched elements aligned with the flow field.
Special treatment of the partition is performed on the wing root juntion and the trailing edge --- into the wake --- to obtain an H-type mesh configuration with anisotropic hexahedra ideal for the strong shear of high Reynolds number simulations.
We then proceed to discretise the far-field using traditional robust tetrahedral meshing tools.
This workflow is made possible by two sets of tools:
CADfix, focused on CAD system, the block partitioning of the near-field and the generation of a linear mesh;
and \textit{NekMesh}, focused on the curving of the high-order mesh and the generation of highly-stretched boundary layer elements.
We demonstrate this approach on a NACA0012 wing attached to a wall and show that a gap between the wake partition and the wall can be inserted to remove the dependency of the partitioning procedure on the local geometry.

\end{abstract}

\newpage

\section*{Nomenclature}
{\renewcommand\arraystretch{1.0}
\noindent\begin{longtable*}{@{}l @{\quad=\quad} l@{}}
1D, 2D, 3D                                   & One-, two-, and three-dimensional, respectively \\
API                                               & Application Programming Interface \\
CAD                                              & Computer-Aided Design \\
CFD                                             & Computational Fluid Dynamics \\
CFI                                               & CADfix interface: its model API \\
CFL                                              & Courant–Friedrichs–Lewy condition \\
NACA                                             & Family of aerofoils developed by the National Advisory Committee for Aeronautics \\
$\Omega$                                   & High-order element \\
$\Omega_\textrm{st}$                  & Reference element \\
$\widetilde{\Omega}_\textrm{st}$ & Subelement of the reference element \\
$\xi$                                              & Parametric coordinates in the reference element \\
$\chi$                                          & Mapping from reference element to high-order element \\
$f$                                              & Affine mapping from reference element to subelement \\
$J_f(\xi)$                                    & Determinant of the Jacobian of the mapping $f$ \\
$P$                                            & Polynomial order \\
$y^+$                                        & Dimensionless wall distance
\end{longtable*}
}

\section{Introduction}

The high-order CFD community currently suffers from the lack of robust mesh generation tools capabale of generating boundary-conforming meshes suited to high-Reynolds number flows of industrial relevance~\cite{Vincent2011,Wang2013}.
Recent efforts have been made using an \textit{a posteriori} approach where a linear mesh is first generated before being curved to represent the boundary representation of the domain accurately.
A review of these techniques can be found in reference~\cite{Turner2017b}.
On the one hand, robust tools exist for the generation of linear unstructured meshes and these have been used successfully by the low-order community for a long time.
\textit{A posteriori} high-order curving of such meshes has however been problematic due to the validity restriction of curvilinear elements.
Highly stretched elements needed for high shear flows are very sensitive to curving therefore requiring curving of the internal mesh in addition to the boundary mesh.
On the other hand, methods based on structured meshing and multi-block partitioning could naturally lend to themselves high-order curvilinear mesh generation.
These meshes are ideal for simple geometries where domain blocks can be mapped to a reference hexahedral block.
This mapping can in turn be used for additional curvilinear refinement~\cite{Moxey2015a,Moxey2015b}.
Such an approach however requires a valid block partitioning, often tedious to obtain on complex indutrial geometries~\cite{Armstrong2015}.

We present an approach that aims at taking advantage of both approaches into a single framework.
We call this approach semi-structured:
a structured mesh is generated in the near-field where requirements of mesh anisotropy, quality and validity are high but also where boundary curvature is important;
then an unstructured mesh is generated in the far-field where those requirements are softened and robust and fast mesh generation is desired.
This approach was successfully applied to generic O-~\cite{Turner2017a} and C-~\cite{Marcon2018} type topologies and we extend it to H-type topologies best described by streamlined bodies.

The method relies on two sets of tools to achieve a final mesh.
On the one hand, we utilise the commercial software CADfix~\cite{ITI-Global2018} for CAD and linear mesh generation.
In the near-field, CADfix uses an approach for block partitioning based on medial objects~\cite{Bucklow2017}.
This allows us to generate coarse boundary layer meshes suitable for high-order curving, which can be later refined to achieve highly-stretched boundary-aligned elements adapted to high-Reynolds number flows.
CADfix also provides a powerful CAD environment capable of CAD healing and manipulation.
This environment is made available through its CFI interface~\cite{Bergemann2018} which \text{NekMesh}, our second set of tools, uses for CAD queries.
From a linear mesh, \textit{NekMesh}, part of the \textit{Nektar++} open-source spectral/\textit{hp} element framework~\cite{Cantwell2015}, generates a curvilinear high-order mesh through projection of high-order nodes onto the boundary representation by querying the CAD model through CFI.

We describe our approach in the following sections.
Section~\ref{sec:semi-structured} first describes the method on a high-level and discusses its specific advantages for H-type topologies.
Section~\ref{sec:cfi} goes on to present CFI, the interface on which \textit{NekMesh} relies heavily on to communicate with CADfix.
Sections~\ref{sec:structured} \&~\ref{sec:unstructured} provide details about the generation of the linear and the high-order meshes in the near- and far-field respectively while Section~\ref{sec:application} presents an example of application of the method.

\section{A semi-structured approach}\label{sec:semi-structured}

We propose a semi-structured approach to high-order curvilinear mesh generation specifically tailored to streamlined bodies.
The typical application of this method is the geeration of a mesh on a wing-fuselage model with a squared-off trailing edge.
The following principles also apply to sharp trailing edges, a special case of squared-off trailing edges where the thickness is zero.
This workflow can also be extended to similar streamlined body configurations.

The meshes generated by our approach are semi-structured:
the near-field and wake are meshed in a structured manner, allowing us to generate highly stretched elements, suitable for high-Reynolds number simulations;
the far field is meshed in an unstructured and robust manner.
The approach takes advantage of both approaches in areas where they are best suited for.
The structured mesh part relies on a block partitioning of the domain based on medial object technology~\cite{Bucklow2017} available in CADfix~\cite{ITI-Global2018}.
A coarse boundary layer mesh can be easily generated and extended into the wake of the streamlined body.
This thick boundary layer, composed of a single element in the thickness, is suitable for high-order mesh generation as it provides enough space for elements to be curved while remaining valid.
This operation is performed in \textit{NekMesh}.
An isoparametric approach~\cite{Moxey2015a} is then employed to refine the coarse boundary layer mesh in the wall-normal direction.
Highly stretched yet valid curvilinear elements can be easily generated this way, following certain restrictions of the mapping~\cite{Moxey2015b}.
The mesh is finally returned to CADfix where the far-field is meshed with pyramids (to interface any quadrilaterals on the skin of the structured near-field mesh) and tetrahedra in a robust unstructured way.

\section{The CFI interface}\label{sec:cfi}

Mesh generation relies on robust CAD systems for reliable node placement both in low-order and high-order approaches.
High-order mesh generation in particular is very sensitive to CAD quality.
Where poor quality CAD can suffice for low-order applications where linear elements would simply go over distortions in the surface definitions, their high-order equivalent is prone to highly distorted elements with no hope of recovery of a valid mesh.
Because CAD software is often focused on design rather than numerical simulation, powerful CAD repair tools are necessary to obtain high quality CAD.
For this reason, CADfix~\cite{ITI-Global2018} is an environment of choice, able to read CAD files in different formats and from different CAD software and to improve the quality of the CAD definition.

\begin{figure}[htbp!]
   \centering
   \includegraphics[width=0.5\textwidth]{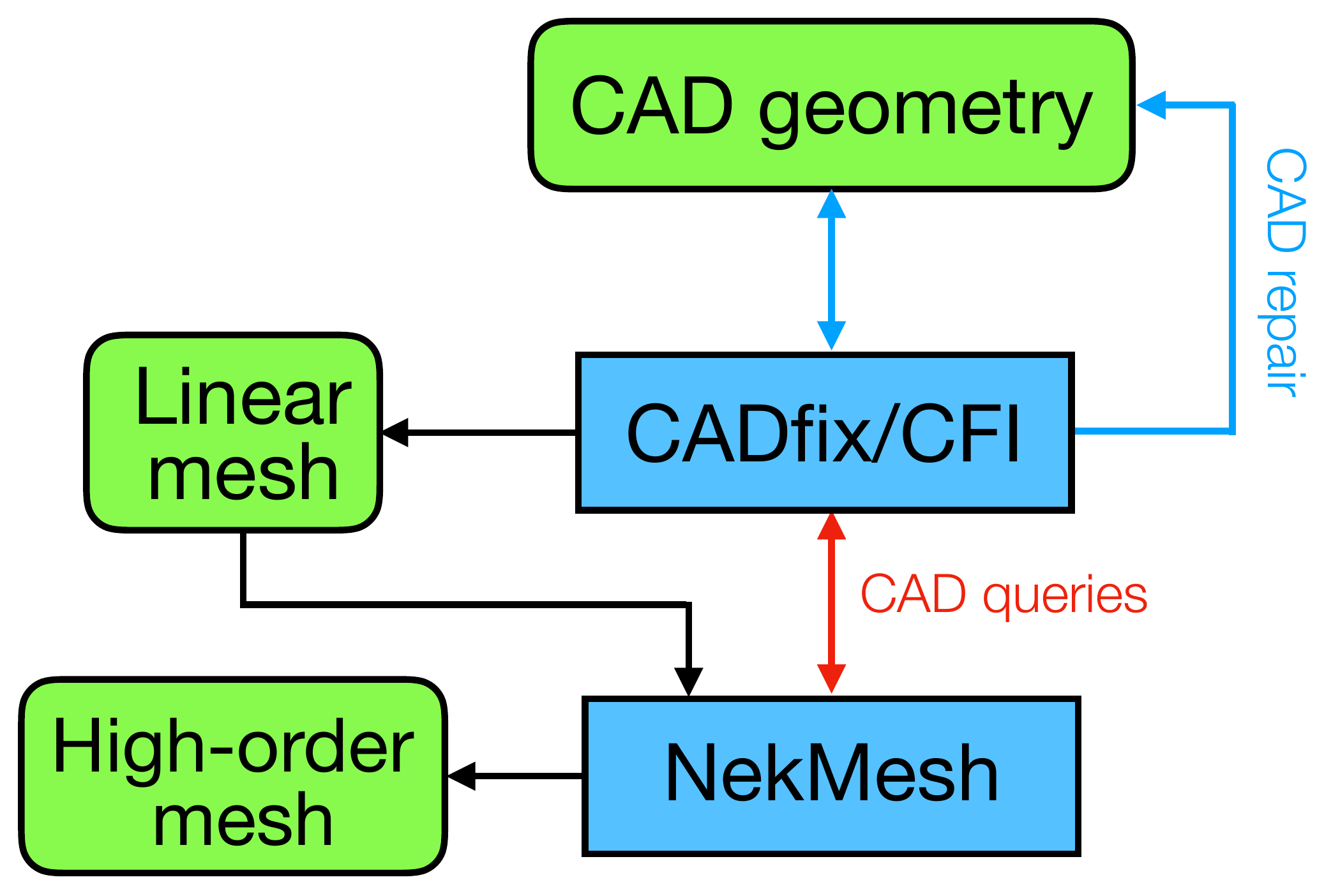}
  \caption{Flowchart of data exchange between \textit{NekMesh} and CADfix through the CFI interface.}
 \label{fig:pipeline}
 \end{figure}

These tools are made available at developer's level through its CFI interface~\cite{Bergemann2018}.
Through polymorphism, \textit{NekMesh} hides the complexity of using a CAD system behind a wrapper.
Initially interfaced with OpenCASCADE~\cite{OpenCascadeSAS2018}, \textit{NekMesh} now supports CFI, not only for access to the CADfix CAD system but also for mesh import and export.
First, the CFI interface is launched by \textit{NekMesh} and it loads the underlying CAD definition into its CAD system.
The linear structured mesh, obtained from the procedure described in Section~\ref{sec:structured}, is then imported from the CADfix database.
An equivalent high-order mesh is generated, following the approach presented in Section~\ref{sec:ho}, at which point extensive use of the CAD system is made through CAD queries.
After splitting of the structured mesh, it is exported into a new database to be processed in CADfix for unstructured meshing of the far-field.
Fig.~\ref{fig:pipeline} illustrates the strong interaction needed for this method to work.

\section{Structured mesh}\label{sec:structured}

Starting from the streamlined body, a coarse structured mesh is generated in the near-field.
It relies on the partitioning of the near-field through medial objects~\cite{Bucklow2017} to generate a coarse boundary layer.
In order to benefit from the quality of structured meshes not only in the boundary layer but also in the wake of the body, these medial objects can be extended downstream and be meshed in a similar fashion.

\subsection{Partitioning based on medial objects}\label{sec:partitioning}

\begin{figure}[htbp!]
	\centering
	\includegraphics[width=0.7\textwidth]{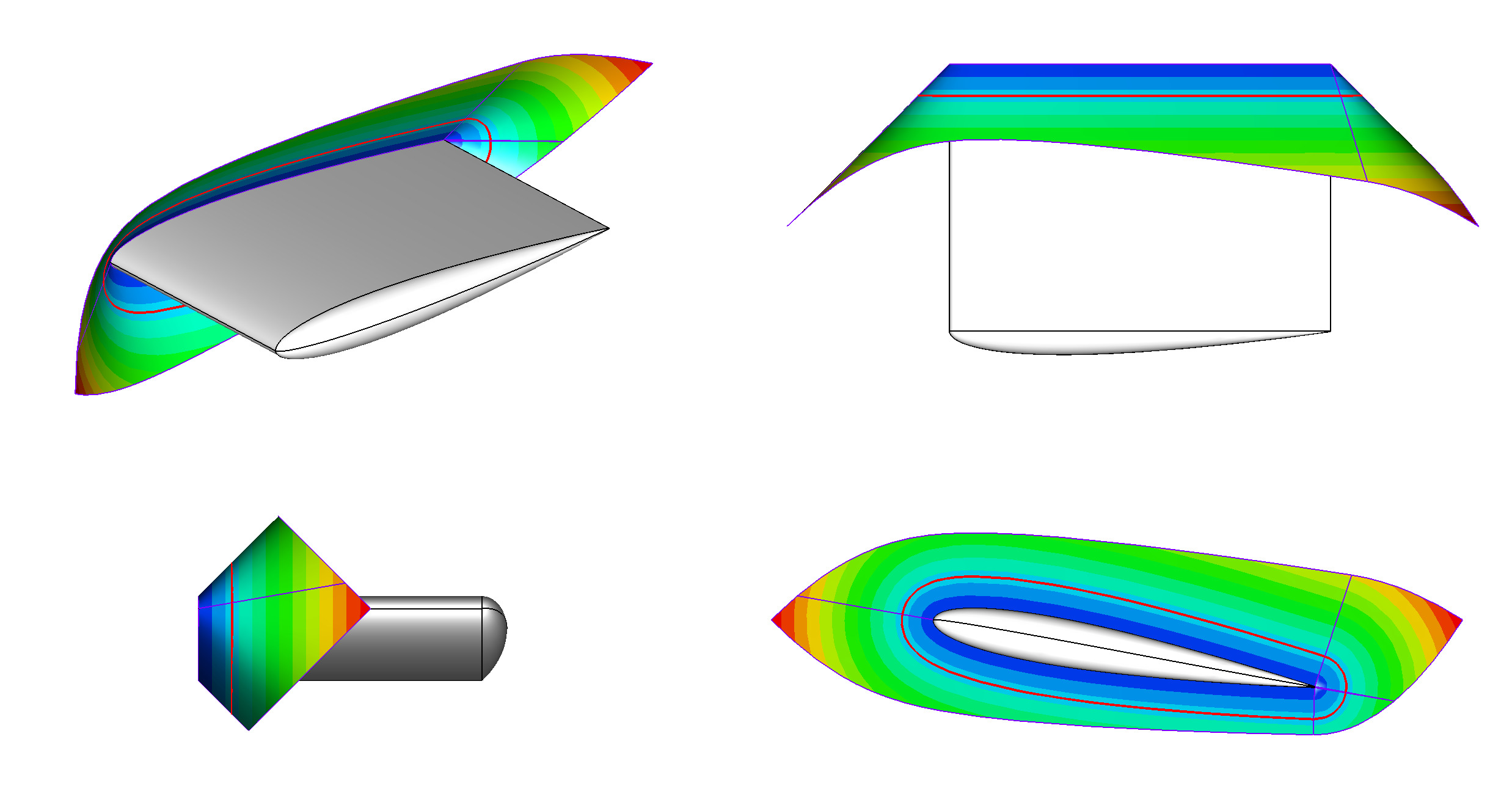} 
	\caption{Example of a medial object for a NACA wing configuration. Medial halos shown in red.}
	\label{fig:medial}
\end{figure}

We use the medial object technology implemented in CADfix~\cite{Bucklow2017,ITI-Global2018}, based on the medial axis first introduced by Blum~\cite{Blum1967} to analyse the topology near the body of interest.
The medial axis is defined as the set of all points inside the domain equidistant to 2 or more boundaries.
CADfix computes these points and assembles them into non-manifold CAD objects, along with extra information such as the connectivities and the medial radius.
These medial objects are used to guide the partitioning of the near-field domain in complex junctions.
See Fig.~\ref{fig:medial}, for an example of three-dimensional medial objects computed around a NACA wing.

\begin{figure}[htbp!]
	\centering
	\includegraphics[width=0.4\textwidth]{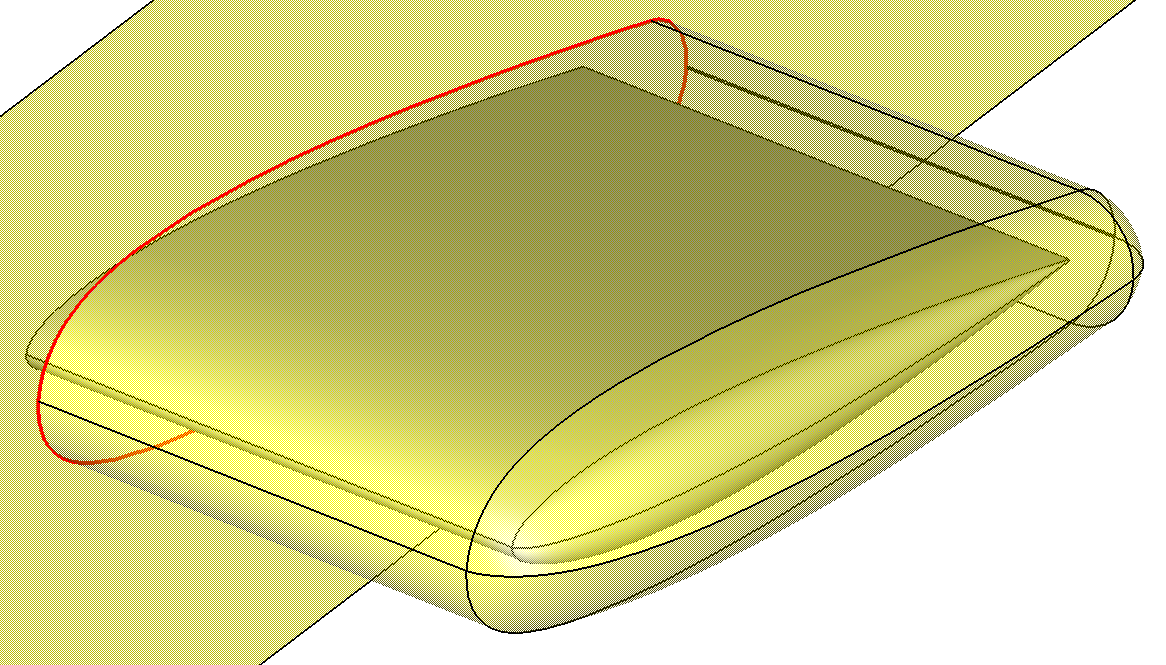}
	\caption{Shell generated around the NACA wing configuration. Medial halos shown in red.}
	\label{fig:shell}
\end{figure}

To generate a valid partition of the near-field, surfaces of the streamlined body are offset and a shell is obtained, see Fig.~\ref{fig:shell}.
This operation essentially divided the domain into a near-field and a far-field partitions.
Issues arise in sharp concave corners and edges where the shell will self-intersect.
These self-intersections form lines called medial halos.
These medial halos, because they are equidistant to at least 2 boundary surfaces, belong both to medial objects and to the shell.
These are visible as red contours on both Fig.~\ref{fig:medial} and Fig.~\ref{fig:shell}.
The medial objects and medial halos together are used to construct the partition around concave areas in contact with the streamlined body.
Once a shell is obtained, the partition is turned into a coarse prismatic layer by sweeping.

\begin{figure}[htbp!]
	\centering
	\begin{tabular}{ccc}
		\includegraphics[width=0.3\textwidth]{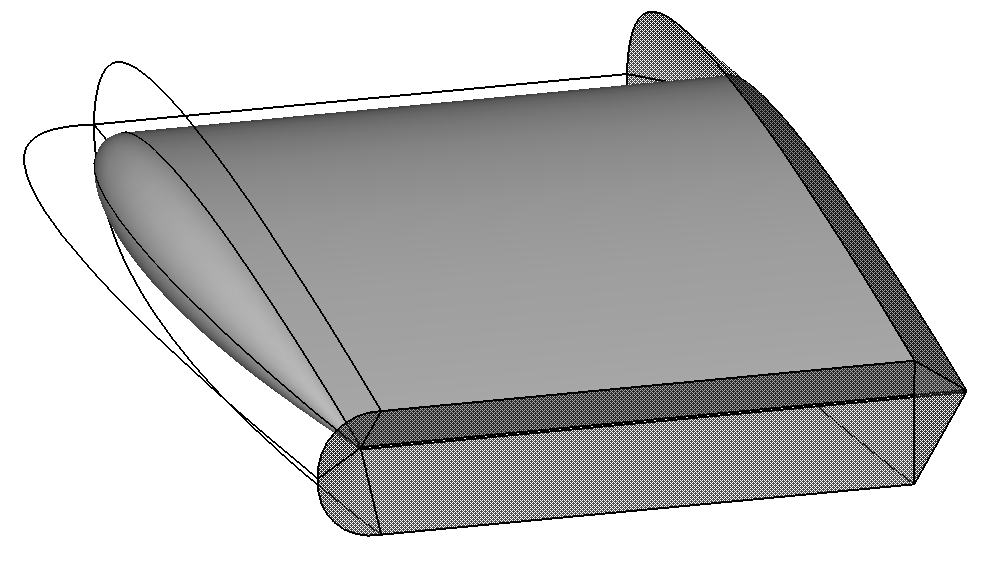}  &
		\includegraphics[width=0.3\textwidth]{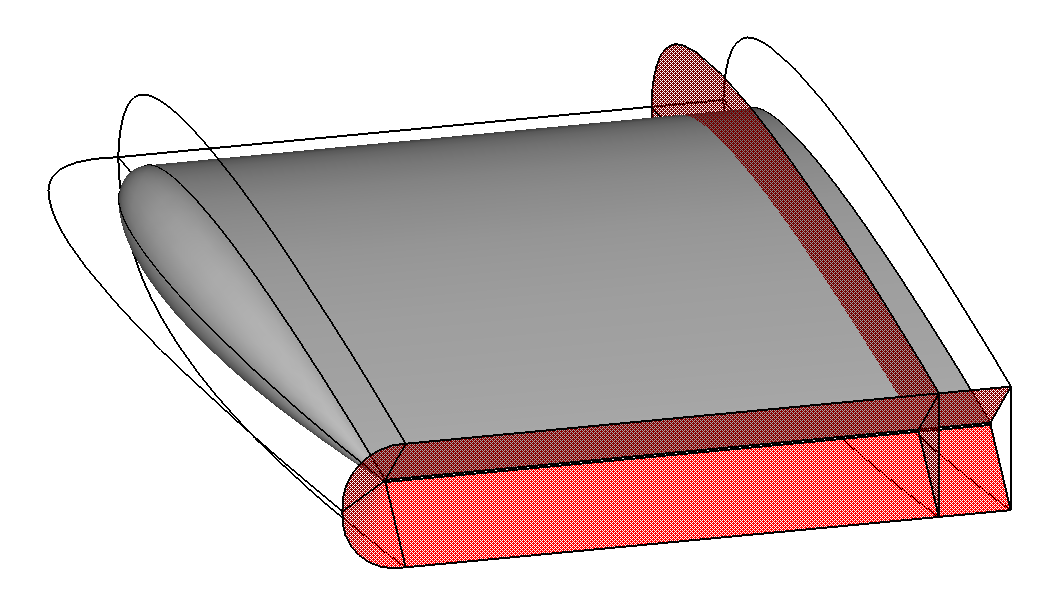}  &
		\includegraphics[width=0.3\textwidth]{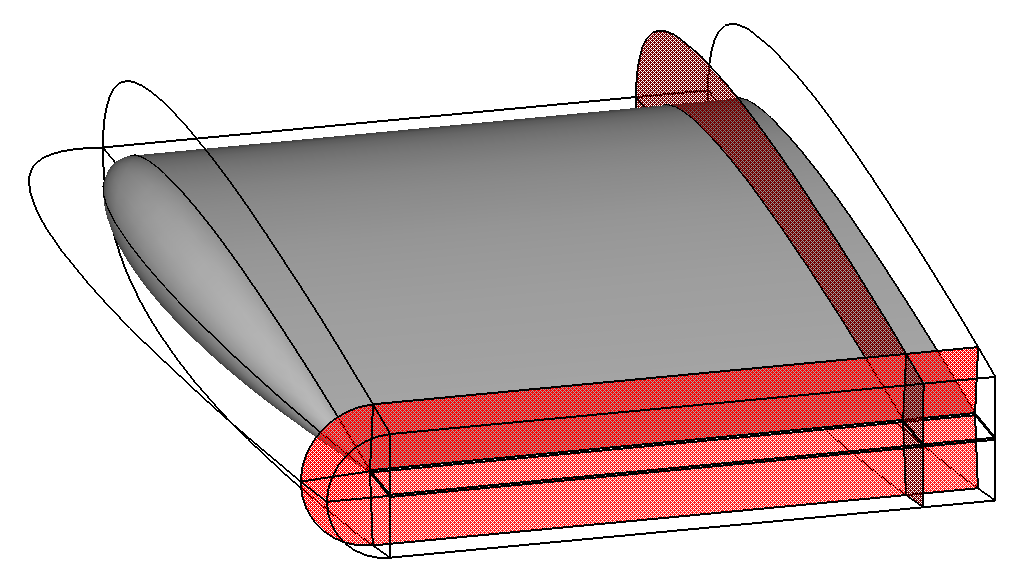} \\
		(a) & (b) & (c) 
	\end{tabular}
	\caption{An illustration of the possible mesh topologies: (a) O-type, (b) C-type and (c) H-type.}
	\label{fig:topology}
\end{figure}

We identify three different mesh topologies, shown in Fig.~\ref{fig:topology} for the NACA wing example.
In the most direct approach, medial objects inside the shell are used to split the boundary layer partition.
This provides us with an O-type topology, see Fig.~\ref{fig:topology}(a), where the whole partition can be mesh with triangular prisms.
This topology is easy to generate but it also creates skewed elements at the wing root junction.
One improvement is the creation of a hexahedral block at that junction.
This is also known as a C-type topology and it is shown in Fig.~\ref{fig:topology}(b).
This hexahedral block can be further split in both wall-normal directions.
The refined mesh at the junction is now suitable for capturing the complex flow patterns generated by the presence of the boundary layers of both the wing and the fuselage.
This topology is well suited for the wing root junction but it is not adequate for the trailing edge.
At the trailing edge, highly stretched elements are obtained, as in the rest of te boundary layer mesh, but their orientation is cross-flow.
This misalignement of trailing edge elements and flow not only restricts the CFL number but it also lacks resolution to resolve shear flow.
A second improvement over the O- and C-type topologies consists in creating a hexahedral block past the trailing edge, known as an H-type topology, as shown in Fig.~\ref{fig:topology}(c).
This allows us to split of the boundary layer mesh, aligned with the flow, improving the quality of the local mesh.

\begin{figure}
	\centering
  \includegraphics[width=0.75\textwidth]{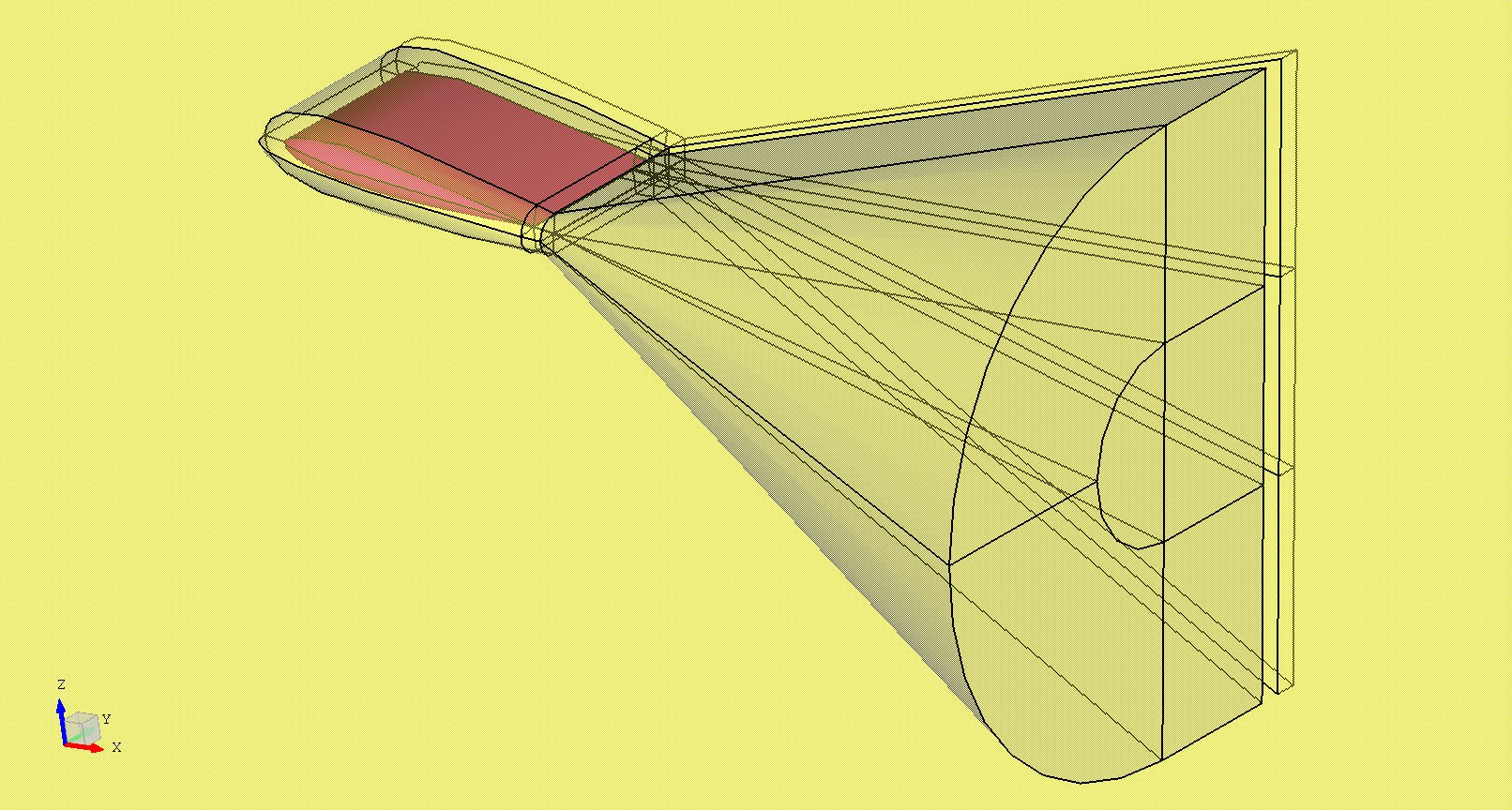} 
	\caption{Example of wake block allowing for increased control of the wake mesh and improved mesh quality.}\label{fig:wake-mesh}
\end{figure}

We present the implementation of this approach with H-type topologies, best suited for streamlined bodies where hexahedral blocks are obtained at the trailing edge, which can be similarly squared-off or sharp.
These trailing edge blocks can be further further extended downstream to create a wake partition.
By doing this, we are able to harness the advantages of structured meshes by generating highly anisotropic hexahedra in the wake of the streamlined body.
This allows the mesh to better capture the shear flow in the same way the structured mesh around the wing is able to resolve the boundary layer flow.
The trailing edge partition can also be manipulated to better suit the need for resolution.
One such manipulation consists in a progressive widening of the partition in the cross-flow direction.
This provides a smoother transition to coarser elements in the unstructured tetrahedral mesh.
Fig.~\ref{fig:wake-mesh} illustrates the capabilities of CADfix in terms of creation and manipulation of wake blocks.

\begin{figure}
	\centering
  \includegraphics[width=0.75\textwidth]{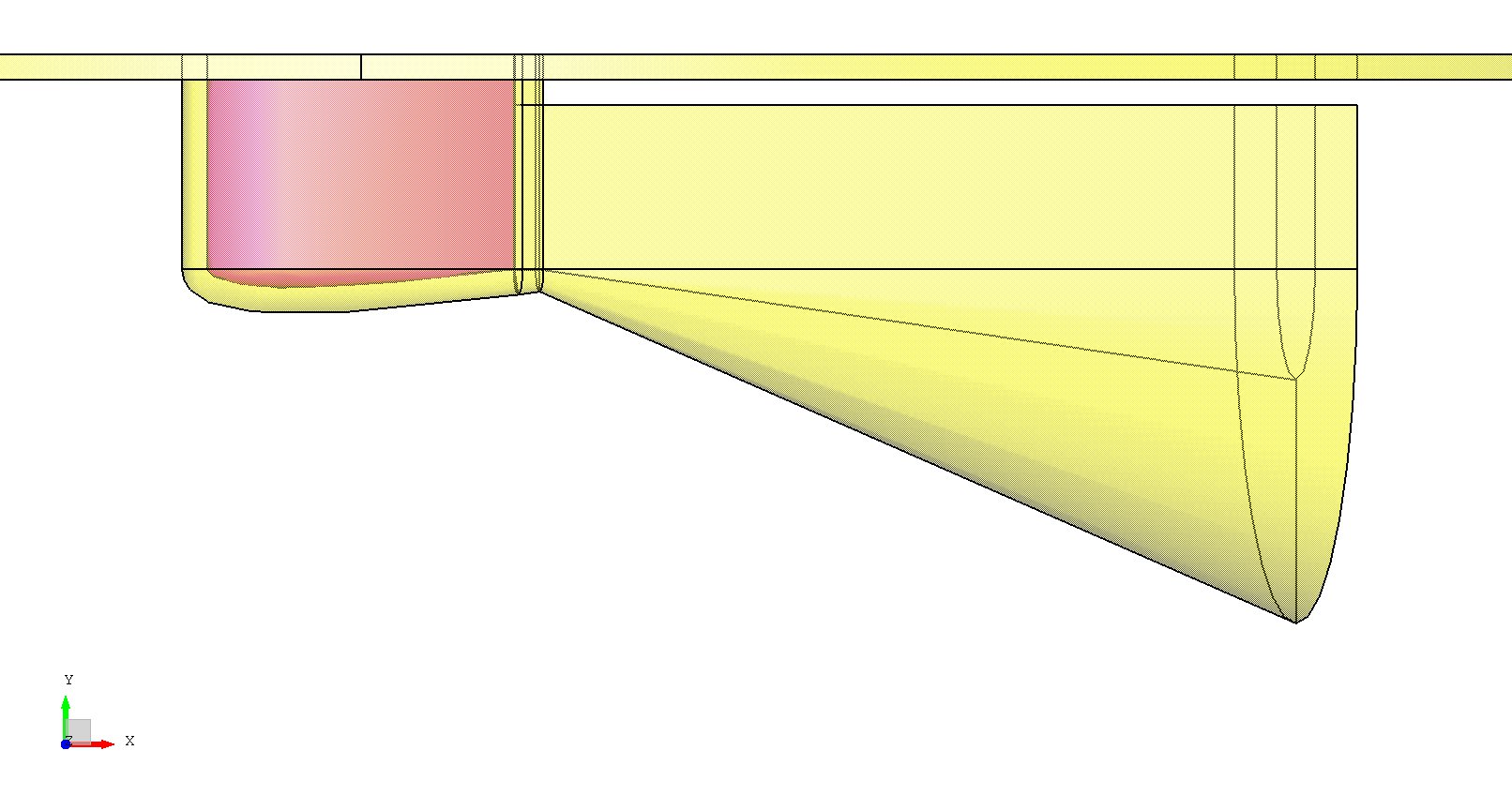} 
	\caption{Example of wake gap with the wall.}\label{fig:wake-gap}
\end{figure}

Difficulty can arise in the extension of the wake partition when the surface to which the wing is fixed isn't plane, e.g.~in a wing-fuselage configuration.
In this situation, the ``wake block root'' not only would be hard to trace along the curved fuselage but it might also expand out of reach in either or both streamwise and cross-flow directions.
For this reason, we introduce a gap between the boundary layer mesh generated at the fuselage or wall and the wake block.
This approach gives us greater flexibility in the generation and manipulation of the wake block partition.
This is illustrated in Fig.~\ref{fig:wake-gap} on a wing clamped to a flat wall.

\subsection{Linear mesh generation}\label{sec:linear-mesh}

The creation of block partitions in the near-field facilitates the generation of a coarse boundary layer mesh.
In each block, a prismatic layer is generated by sweeping from the boundary CAD throughout the near-field partition.
A bottom-up mesh generation approach is actually used to retain conformality between all partitions.
Lines are meshed first, then boundary surfaces by Delaunay triangulation and finally volume blocks by sweeping of the surface meshes.
While a multi-layer volume mesh would usually be created when generating a linear mesh, the difficulty of curving such highly stretched elements makes this practice undesirable for high-order meshes.
A coarse mesh with a single element in the thickness is generated instead., giving more room for valid high-order curving.

\begin{figure}
   \centering
     \includegraphics[width=0.6\textwidth]{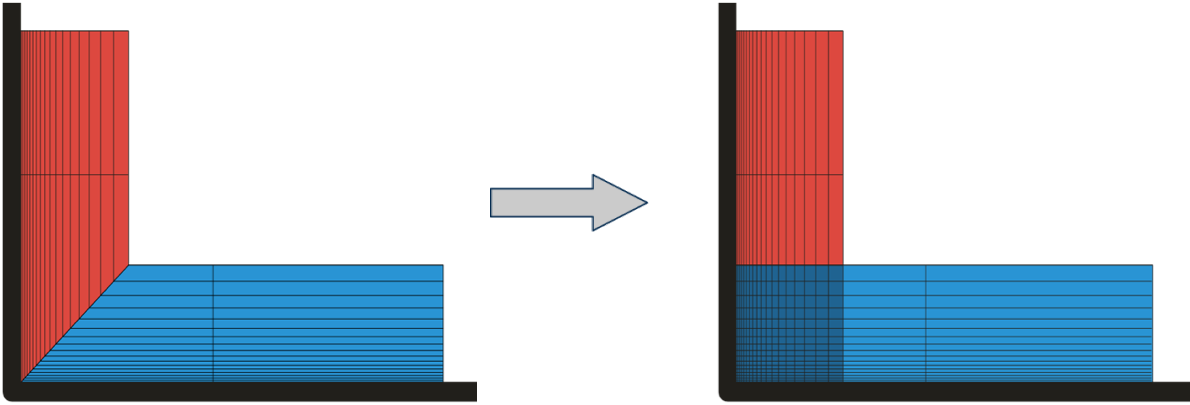}   
   \caption{The standard prismatic partition of O-type topology (left) vs the hexahedral partition at the wing root junction of a C-/H-type topology.}
   \label{fig:rotor67-1}
\end{figure}

Not all block partitions in an H-type mesh topology are filled with triangular prisms though.
The hexahedral blocks at the wing root junction as well as in the wake cannot be meshed with prisms due their interface with the side of prismatic partitions.
Instead, these hexahedral blocks are meshed with a single hexahedron in the thickness of the boundary layer.
At the wing root junction, the block partition is connected to two distinct boundary surfaces and therefore belongs to two different boundary layers.
A single hexahedron is thus generated in each direction.
Fig.~\ref{fig:rotor67-1} illustrates the gain in quality of using a C-/H-type topology at the wing root junction over an O-type approach.
In the wake, on the other hand, a split can be made in the streamise direction.
Indeed, the wake blocks are not in contact with any boundary surface other than the negligible trailing edge.
Quads obtained on the rear interface of the wing's prismatic layer can be swept downstream in a multi-layer fashion.
  
\subsection{High-order mesh generation}\label{sec:ho}

This coarse structured mesh makes the \textit{a posteriori} generation of a high-order mesh less prone to invalid elements.
The thickness of the coarse layer offers enough space to accommodate boundary curvature and avoid self-intersecting elements.
The projection to high-order, i.e.~the curving of elements, follows a bottom-up approach presented in~\cite{Sherwin2002}.
High-order nodes are placed along the edges of the linear elements then projected onto their parent CAD object.
The same then goes for faces.
Optimisation of the location of the nodes after projection typically provides a large improvement in high-order element quality.
This is especially important for edge and face nodes projected onto surfaces.
This optimisation consists in the minimisation of the deformation of a spring system around those nodes.
First edge nodes are optimised while end nodes (existing from the linear mesh) are fixed.
After optimisation, the high-order nodes are expected to approximately lie on the geodesic between the two end points.
We then go on to optimise the location of face nodes by fixing all linear and high-order edge nodes.
A more complex spring system is set up where each face node is connected to its neighbouring nodes and the deformation is minimised again.

\begin{figure}
  \centering
     \includegraphics[width=0.98\textwidth]{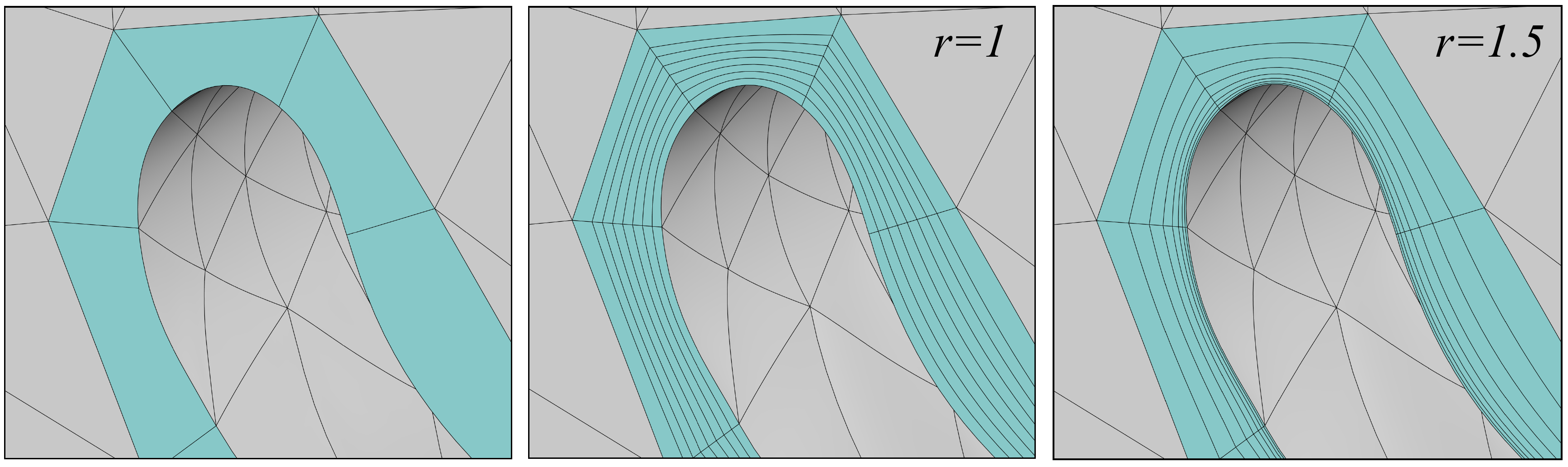}   
     \caption{Illustration of the isoparametric splitting approach: from a coarse boundary layer mesh (left) to a split mesh with uniform (centre) and progressive (right) distribution of elements in the wall normal direction.}
  \label{fig:bl_mesh}
\end{figure}

A high-order curvilinear near-field mesh is obtained, consisting of a single layer of curved triangular prisms and hexahedra.
This boundary layer is however too coarse for high-Reynolds number CFD simulations where gradients near the wall in the cross-flow are large. 
Splitting of the coarse boundary layer elements is necessary to achieve highly stretched elements able to provide enough resolution in the wall normal direction without restricting the CFL number in the streamwise direction.
This wall normal refinement is achieved through the isoparametric splitting approach presented in~\cite{Moxey2015a}.
Starting from the mapping \( \chi \) between a reference element \( \Omega_{\text{st}} \) and the physical space element \( \Omega \), subdivisions are made by adding points along the height of the prisms and hexahedra.
Because all nodes are introduced with respect to mapping \( \chi \), the split mesh retains the validity and quality of the coarse mesh.
The number and height of subdivisions can be specified by the user according to the resolution needs.
This allows us to reach aspect ratios virtually unachievable through traditional \textit{a posteriori} mesh curving techniques.
This technique is illustrated in Fig.~\ref{fig:bl_mesh} where a coarse prismatic layer (left) is split into fine multi-layer meshes with different growth ratios (centre and right).

\begin{figure}[htbp!]
  \begin{center}
    \includegraphics[width=0.7\textwidth]{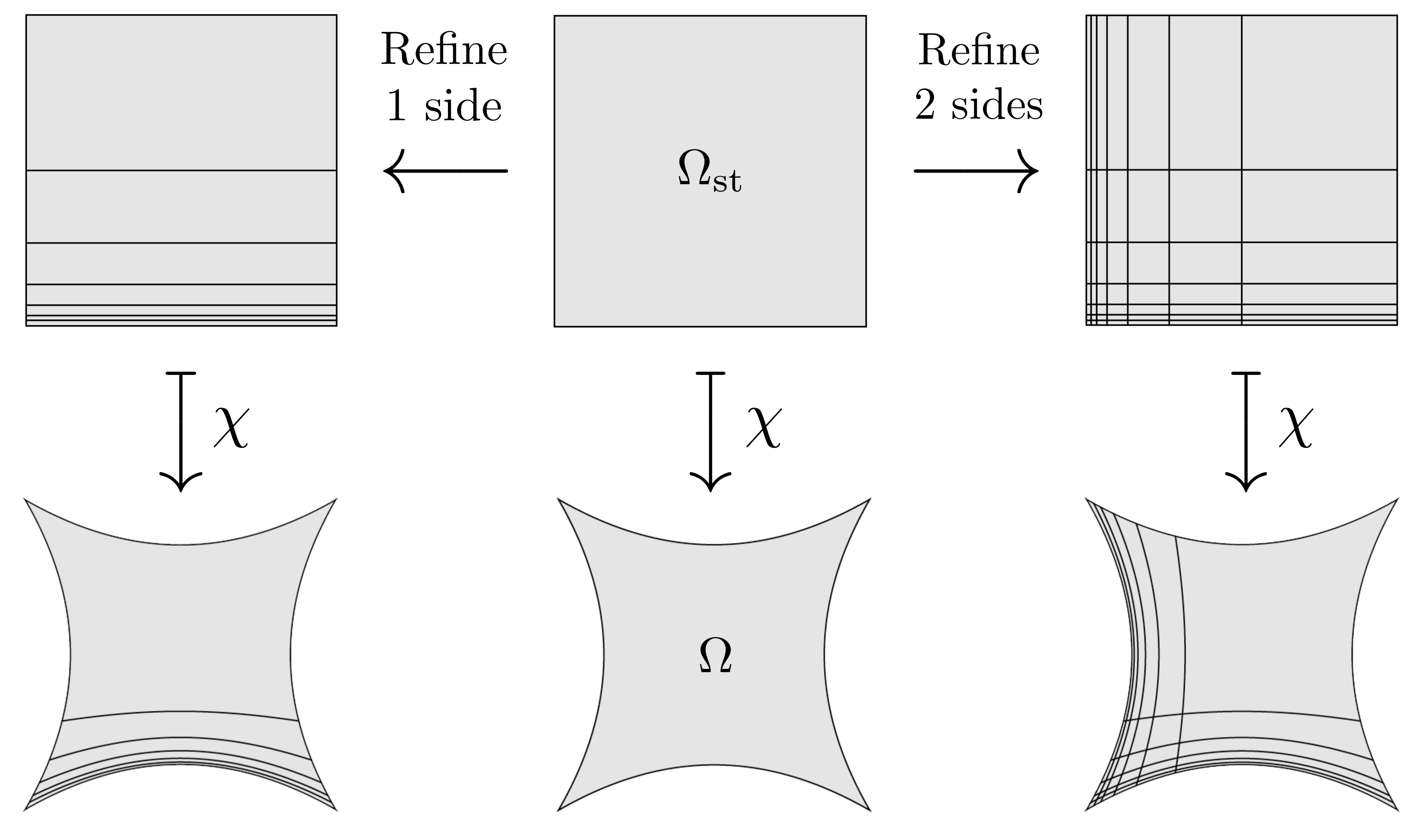}
  \end{center}
  \caption{Extension the isoparametric splitting approach to bi-direction splitting.}
  \label{fig:bl-refined}
\end{figure}

While the isoparametric splitting approach was originally developed for O-type topologies, special care must be given to the hexahedral blocks located at wing root junctions.
These hexahedra belong to two different boundary layers and need therefore to be split in both directions sequentially.
A visual demonstration is shown in Fig.~\ref{fig:bl-refined} where the initial version of the isoparametric splitting in one direction (left) was extended to bi-directional splitting (right).
This natural evolution of the isoparametric splitting produces high quality elements at wing root junctions with added resolution in both directions of interest, able to capture the high wall normal gradients produced by the two neighbouring boundaries.

\section{Unstructured mesh}\label{sec:unstructured}

We obtain a structured high-order curvilinear mesh in the near-field with high aspect ratio elements near the walls, where crucial phenomena in high-Reynolds number simulations require important resolution.
As we move away from the wall, the flow field becomes smoother and coarser mostly isotropic elements are preferred.
While traditional structured meshing techniques require a tedious block decomposition of the whole domain, we resort to an unstructured mesh generation approach.
In the far-field, highly anisotropic elements are not required and there exists no curvature prone to render high-order elements invalid.
We resort to easy-to-use and robust tetrahedral meshers for this part of the domain.
The high-order mesh is first exported to a CADfix database file through the CFI interface.
CADfix is then able to extract the skin mesh, i.e.~the surface mesh of the shell generated during the block partitioning, and use it as an input for the tetrahedral mesher.
Unlike in the O- and C-type topologies, the near-field mesh obtained in the H-type configuration is not only composed of triangles but also of (split) quadrilaterals.
These quadrilaterals are located all around the wake block and require a pyramidal mesh layer to transition from the hexahedral block mesh to the unstructured tetrahedral mesh.

\section{Example of application}\label{sec:application}

\begin{figure}[htbp!]
   \begin{center}
      \includegraphics[width=0.49\textwidth]{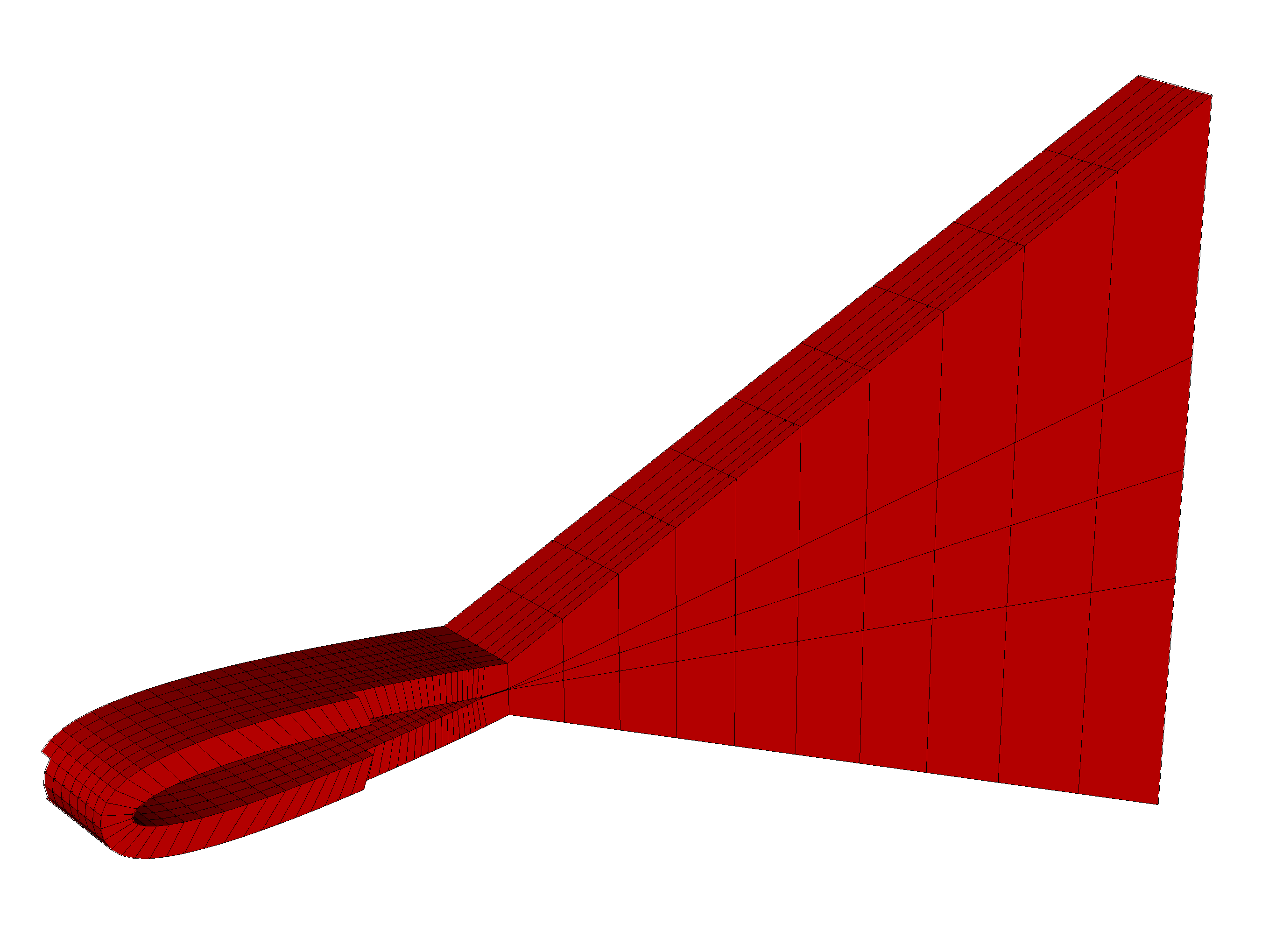}
      \includegraphics[width=0.49\textwidth]{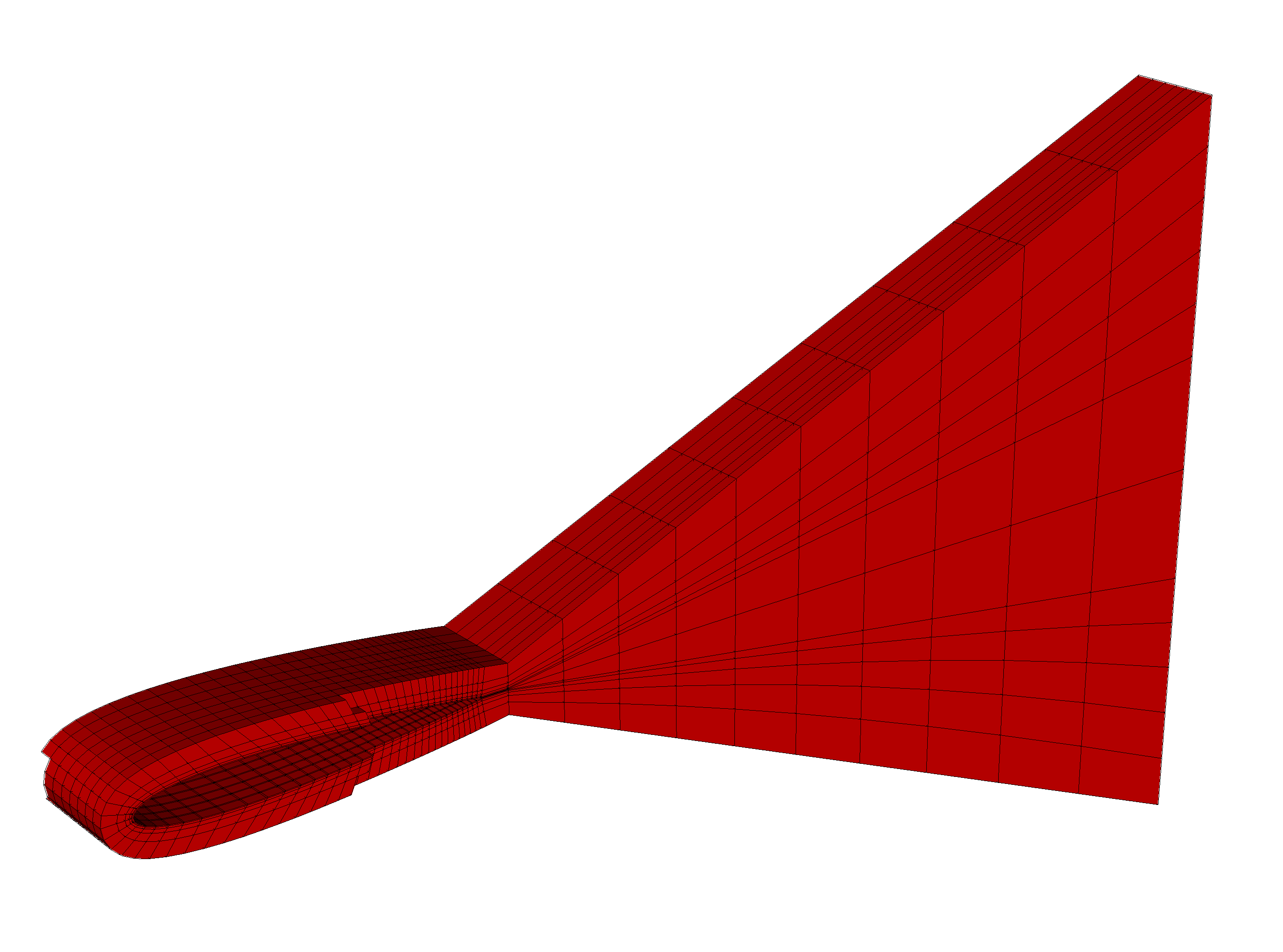}
   \end{center}
   \caption{Coarse (left) and split (right) near-field meshes obtained on the NACA wing domain with H-type topology.}
   \label{fig:naca-span}
 \end{figure}

We illustrate this semi-structured approach to high-order mesh generation around a streamlined body on a traditional NACA0012 wing.
This wing has a squared-off trailing edge of finite thickness and is attached to a flat wall.
The near-field is partitioned using an H-type topology as presented in Section~\ref{sec:partitioning} and the medial objects are shown in Figs~\ref{fig:wake-mesh} \&~\ref{fig:wake-gap} with a gap between the wake block and the wall.
The near-field is then discretised with a linear mesh, then transferred to \textit{NekMesh} for high-order meshing and finally split, following the procedures described in Sections~\ref{sec:linear-mesh} \&~\ref{sec:ho}.
The structured mesh on a portion of span before and after splitting is shown in Fig.~\ref{fig:naca-span}.
A progression ratio of \(2.0\) and \(5\) elements in the thickness were used for the splitting in the boundary layer of the wing.
The progression then linearly varies with distance from the trailing edge until reaching \(1.0\) (uniform distribution of elements) at the far end of the wake.

\section{Conclusions}

The adoption of a semi-structured approach, where a multi-block partition is only used in the near-field region and the rest of the computational domain is discretised via an unstructured mesh, permits the generation of high-quality, highly stretched meshes in the boundary layer and wake regions.
Since we are not required to obtain a partition in cubes of the full domain, this eliminates one of the bottlenecks that often prevent the use of a full multi-block for some complex geometries.
Further, the generation of a coarse structured mesh in the near-field where a split into layers along the normal direction via the isoparametric approach facilitates the generation of valid high-order meshes with large stretching that inherit their quality from the coarse structured mesh.
We have adopted an H-type topology of the mesh in the near-field to incorporate wakes shed from wing boundary layers.
To increase the flexibility of the method, we allow quadrilateral elements on the discretisation of the surface.
The discretised near-field region is a hybrid mesh of triangular pyramids and hexahedra and the outer mesh incorporates quadrilateral pyramids in the transition to the near-field mesh.
The proof-of-concept example presented illustrates the ability of the method to generate this type of meshes and also the ease with which the resolution within the boundary layer can be controlled to achieve values of $y^+$ of order 1 and below for the location of the first node away from the surface.

\section*{Acknowledgments}

This project has received funding from the European Union's Horizon 2020 research and innovation programme under the Marie Sk\l{}odowska-Curie grant agreement No 675008.

\bibliography{refs}

\end{document}